\documentclass[amsmath,amssymb,reprint,aps,pra,floatfix]{revtex4-2}
\usepackage{graphicx}
\usepackage{lineno}

\usepackage[utf8]{inputenc}
\usepackage{siunitx}
\usepackage[german, english]{babel}

\begin{document}

\title{Stabilizing the free spectral range of a large ring laser}

\author{Jannik Zenner}
  \affiliation{Physikalisches Institut, Universität Bonn, Nussallee 12, 53115 Bonn, Germany}
\author{Karl Ulrich Schreiber}
  \affiliation{Research Unit Satellite Geodesy, Technical University of Munich, 80333 Munich, Germany}
  \affiliation{School of Physical Sciences, University of Canterbury, Christchurch 8140, New Zealand}
\author{Simon Stellmer}
  \email{stellmer@uni-bonn.de}
  \affiliation{Physikalisches Institut, Universität Bonn, Nussallee 12, 53115 Bonn, Germany}

\date{\today}

\begin{abstract}

A ring laser is defined by its perimeter, which directly enters the conversion factor between measured Sagnac frequency and the actual rotation rate. Large ring lasers employed in geodesy and fundamental physics require stability of the perimeter at or below the parts-per-billion level. We present two complementary approaches to actively control the perimeter length of such ring lasers, reaching a relative length stability of $4\times 10^{-10}$. One of these approaches is based on a phase detection between the beat of two resonances of different longitudinal mode index and a stable local oscillator. The other approach employs a highly stable wavelength meter to measure the absolute frequency of the laser light. These methods can readily be implemented and bring the stability of heterolithic devices on par with monolithic designs.

\end{abstract}

\maketitle

\section{Introduction}

The winter 2024/25 marks the centennial of the famous Michelson-Gale-Pearson experiment \cite{Michelson1925a,Michelson1925b}. On an open field near Chicago, A.~A.~Michelson and his co-workers had installed a large rectangular interferometer, with arm lengths measuring 612 by 339 meters. Two light fields propagating in opposite directions along the contour will acquire a phase shift proportional to the rotation rate of the setup, known today by one of the pioneers in optical rotation measurements, G.~Sagnac \cite{Sagnac1913}. The crew of 1924 observed a fringe shift of $0.270\,\lambda$ when comparing the interference pattern of the large ring with a zero-area reference setup. This observation agreed to within $2\%$ with the calculated value from the Earth rotation rate. This hallmark experiment was, although unintended, the first measurement of Earth rotation based on the Sagnac effect, and one of the first tests of special relativity in the limit of small velocities.

Today, various experiments around the world are targeted at highly sensitive rotation measurements using large, meter-sized ring resonators in the fields of geodesy \cite{Schreiber2013,Gebauer2020}, seismology \cite{Igel2005, Igel2011}, and fundamental physics \cite{Stedman1997, Bosi2011}. Some of these instruments are contemporary versions of the 1924 setup, so-called passive ring resonators with an external laser feeding two counter-propagating resonator modes \cite{Ezekiel1977,Liu2019,Korth2016}. Alternatively, the laser medium, usually a helium-neon gas, may be placed inside the ring resonator, forming a ring laser. To our knowledge, only two of these instruments are quasi-monolithic structures with very small mechanical instabilities: the C-II ring in Canterbury (New Zealand), which was the first ring laser to measure low frequency geophysical signals beyond the rotation of Earth \cite{Schreiber2003}, and the G~ring in Wettzell (Germany), which currently is the frontrunner in terms of sensitivity and stability \cite{Schreiber2023}. All heterolithic devices will benefit from an active stabilization of the resonator perimeter. Such approaches have been demonstrated for passive ring resonators \cite{Zhang2020}, but not for ring lasers.

Here, we present two approaches that allow stabilization of the perimeter of a ring laser to a few nanometer, equivalent to a relative instability in the $10^{-10}$ range.

Consider a ring-shaped optical resonator with perimeter $P$ and enclosed area $A$, rotating at a rate $\Omega$. The Sagnac frequency of two counter-propagating modes of the same longitudinal mode index $m$ reads
\begin{align}
    \label{align:sag}
    \delta f = \frac{4 A}{\lambda P} \Omega \sin\theta,
\end{align}
where $\lambda$ is the wavelength of the light and $\theta$ is the projection angle of the surface vector onto the rotation vector. The dimensionless pre-factor $4A/\lambda P$ is called the scale factor. Another important quantity is the free spectral range (FSR), which is the frequency splitting between adjacent longitudinal modes of the the resonator, $f_{\text{FSR}} = c/P$. The FSR is directly proportional to the ring perimeter $P$, and we will present two approaches to stabilize the perimeter through stabilization of the FSR.

\section{Setup}

\begin{figure}[t]
    \includegraphics[width=\columnwidth]{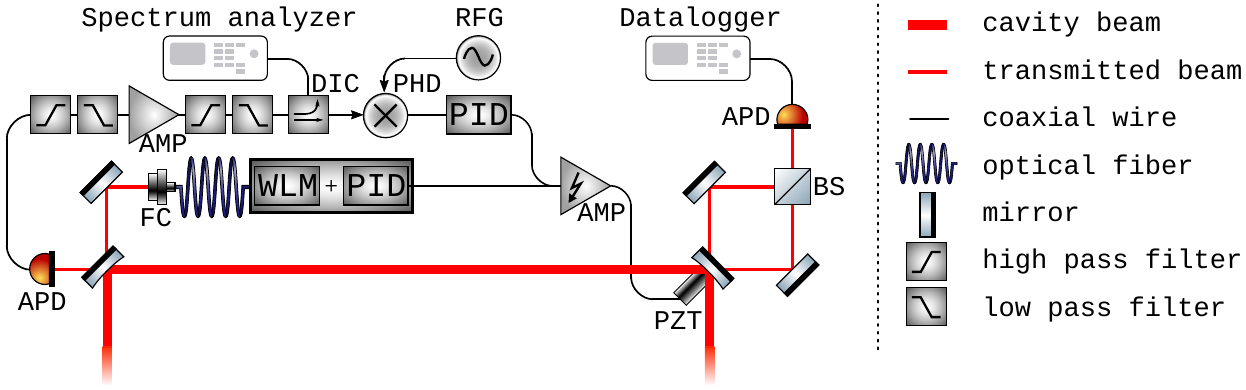}
    \caption{\label{fig:setup} Schematic view of the setup to stabilize the perimeter of the ring laser cavity using either the \textit{absolute~frequency~lock} or \textit{FSR~phase~lock} method (left-hand side corner) and to measure the Sagnac frequency (right-hand side corner). Abbreviations: AMP \textit{radio frequency amplifier}, APD \textit{avalanche photodetector}, BS \textit{beamsplitter}, DIC \textit{directional coupler}, FC \textit{fiber coupler}, PHD \textit{phase detector}, PID \textit{proportional-integral-derivative controller}, PZT \textit{piezo actuator}, RFG \textit{radio frequency generator}, WLM \textit{wavelength meter}.}
\end{figure}

The setup is based on a square ring laser with a side length of about \SI{3.5}{\meter}, the details of which have been described elsewhere \cite{Schreiber2009,Schreiber2006}. The resonator is formed by four mirrors with a radius of curvature of \SI{4}{\meter} and has a finesse of about 40,000. The ring laser usually operates at an absolute frequency of \SI{473612110 \pm 2}{\mega \hertz}, the free spectral range is \SI{21423199.0 \pm 0.1}{\hertz} and the inferred mode index, corrected for various small systematics detailed in Ref.~\cite{Hurst2017}, is $m=22107440.0(0.4)$. The measured Sagnac frequency of $\delta f=$ \SI{311.65}{\hertz} is in very good agreement with expectations from the known Earth rotation rate $\Omega_E$ and the latitude of the ring laser, $\theta' = 50^{\circ}$ 43' 41.9'' N.

The technical setup is shown in Fig.~\ref{fig:setup}. At one of the ring cavity mirrors, a 3.5-ppm fraction of the two counter-propagating light fields is coupled out of the cavity, passes through anti-reflection coated vacuum windows, and is superimposed on a non-polarizing 50:50 beamsplitter. The two path lengths between cavity mirror and beamsplitter are equal to within a millimeter. The signal is measured with an avalanche photodetector and digitized with a data logger. The modulation of the Sagnac beat signal has an amplitude greater than 95\%.

The cavity mirror holder of that same corner is fixed to a multilayer stack piezo actuator with a total motion range of \SI{89}{\micro \meter} at a variation of \SI{0.6}{\micro \meter \per \volt} oriented orthogonal to the mirror surface, allowing for a perimeter variation of up to $\Delta P = $ \SI{126}{\micro \meter} and a possible shift of the FSR of up to $\Delta f_{\text{FSR}} = $ \SI{193}{\hertz}. The load-free resonance frequency of \SI{3}{\kilo \hertz} should be reduced to about \SI{150}{\hertz} due to the large inert mass of about \SI{10}{\kilogram} of the mirror mount chamber. Experimentally, it is observed that the movement bandwidth is already limited to about \SI{17}{\hertz}, due to further attenuation of the movement caused by the bellows of connected vacuum pipes and friction of the stage, the piezo is connected to.

To stabilize the perimeter of the ring laser cavity, the piezo is steered according to one of two parameters detected at the left-hand side of Fig.~\ref{fig:setup}. The light is either coupled into a fiber, which is then used to perform an \textit{absolute~frequency~lock}, or it is detected by a \SI{400}{\mega \hertz} bandwidth avalanche photodetector (APD) for a \textit{phase~lock} of the FSR. Drifts of the FSR occur on the timescale of a minute, thus the locking bandwidth can be very slow. It is limited by the bandwidth of the piezo and, for the absolute frequency lock, by the update rate of \SI{0.25}{\hertz}.

\subsection{Absolute frequency lock}

About \SI{40}{\nano \watt} of the light transmitted through one of the mirrors is coupled into a \textit{HighFinesse WS8} wavelength meter (WLM) to measure the absolute frequency of the light field $f_{\text{L}}$. At this power, \SI{4}{\second} of integration time are needed per data point, limiting the locking bandwidth to \SI{0.25}{\hertz}. With a resolution of \SI{0.1}{\mega \hertz} and an accuracy of about \SI{2}{\mega \hertz}.
The reading of the absolute frequency from the wavemeter is limited by drifts on timescales longer than a few minutes, likely caused by pressure and temperature changes in the laboratory. We quantify these drifts through comparison with an ultrastable reference laser. Typical drift rates are \SI{1}{\mega\hertz / \hour} \cite{Zhang2013}. The measured value is transferred to a digital proportional-integral-derivative (PID) controller module. A high voltage amplifier drives the piezo actuator and amplifies the PID output by a factor of ten. 

\subsection{FSR phase lock}

The width of the gain profile of the helium-neon laser is about \SI{1.8}{\giga \hertz} and allows the laser to run on a multitude of longitudinal modes, each of them spaced by one FSR (about \SI{21.4}{\mega \hertz}). The number of active modes increases with the driving power, and the driving power can be adjusted such that the helium-neon plasma lases only on one single mode. Here, we set the driving power to a slightly higher value, such that small additional modes appear. While modes close to the main carrier are completely suppressed, a secondary mode at $4 \cdot f_{\text{FSR}} = \SI{85.6928}{\mega \hertz}$ appears. All higher modes are at least \SI{10}{\decibel} weaker. 
This narrow operation regime where the Sagnac beat is not yet perturbed due to multimode competition is only present at \SI{110}{\nano \watt} to \SI{120}{\nano \watt} of available light intensity that is transmitted out of the cavity. Detecting this on a high-bandwidth APD, the beat signal at $4 \cdot f_{\text{FSR}}$ has a stable level of \SI{-62}{\decibel m} and requires significant filtering and amplification. Good results are achieved by sharp low and high pass filtering of the signal, subsequent amplification with a combination of two amplifiers, and a second set of filters. The combination of a high pass filter with a \SI{1}{\decibel} pass band above \SI{90}{\mega \hertz} and a low pass filter with a \SI{1}{\decibel} pass band below \SI{81}{\mega \hertz} lead to a combined attenuation of \SI{2.5}{\decibel} at the signal frequency, but result in a drastic improvement of the signal-to-noise ratio (SNR) by several orders of magnitudes, resulting in a SNR of better than \SI{25}{\decibel}. An extremely low noise figure (NF) amplification is done by a first amplifier (NF: \SI{1}{\decibel}, gain: \SI{25}{\decibel}) followed by a high-gain amplifier (NF:  \SI{6}{\decibel}, gain: \SI{34}{\decibel}). The resulting signal, together with a highly stable RF signal derived from a synthesizer that is referenced to a \textit{Stanford Research Systems FS725} atomic clock, are input into a phase detector. The output of that phase detector is a \SI{\pm 0.8}{\volt} DC voltage proportional to the phase difference of both inputs with a great figure of merit of 143, that therefore outputs a near linear error signal of \SI{0.8}{\volt \per \radian}, which sets the sensitivity of this scheme. This voltage is stabilized by a PID controller (resolution: \SI{2}{\milli\volt}) to better than \SI{100}{\milli\volt} by steering the piezo actuator via a high voltage amplifier (\SI{0.3}{\milli\volt} ripple), thereby stabilizing the FSR to better than \SI{1}{\hertz}. Recalling that $f_{FSR} = c/P$, it is clear that any lock of the FSR also locks the perimeter $P$ and the wavelength $\lambda$ and thus stabilizes the scale factor.\\
The bandwidth of this locking scheme is limited to below about \SI{17}{\hertz} by the relatively large inert mass of about \SI{10}{\kilogram} moved by the piezo. A more lightweight setup, where only the mirror is moved (three orders of magnitude lighter) by a ring piezo instead of the entire stainless steel mirror mount chamber could increase the bandwidth further into the acoustic regime to above \SI{2}{\kilo \hertz}, at possibly the cost of worse access to the transmitted beams.

\section{Results}

\begin{figure} [t]
    \includegraphics[width=\columnwidth]{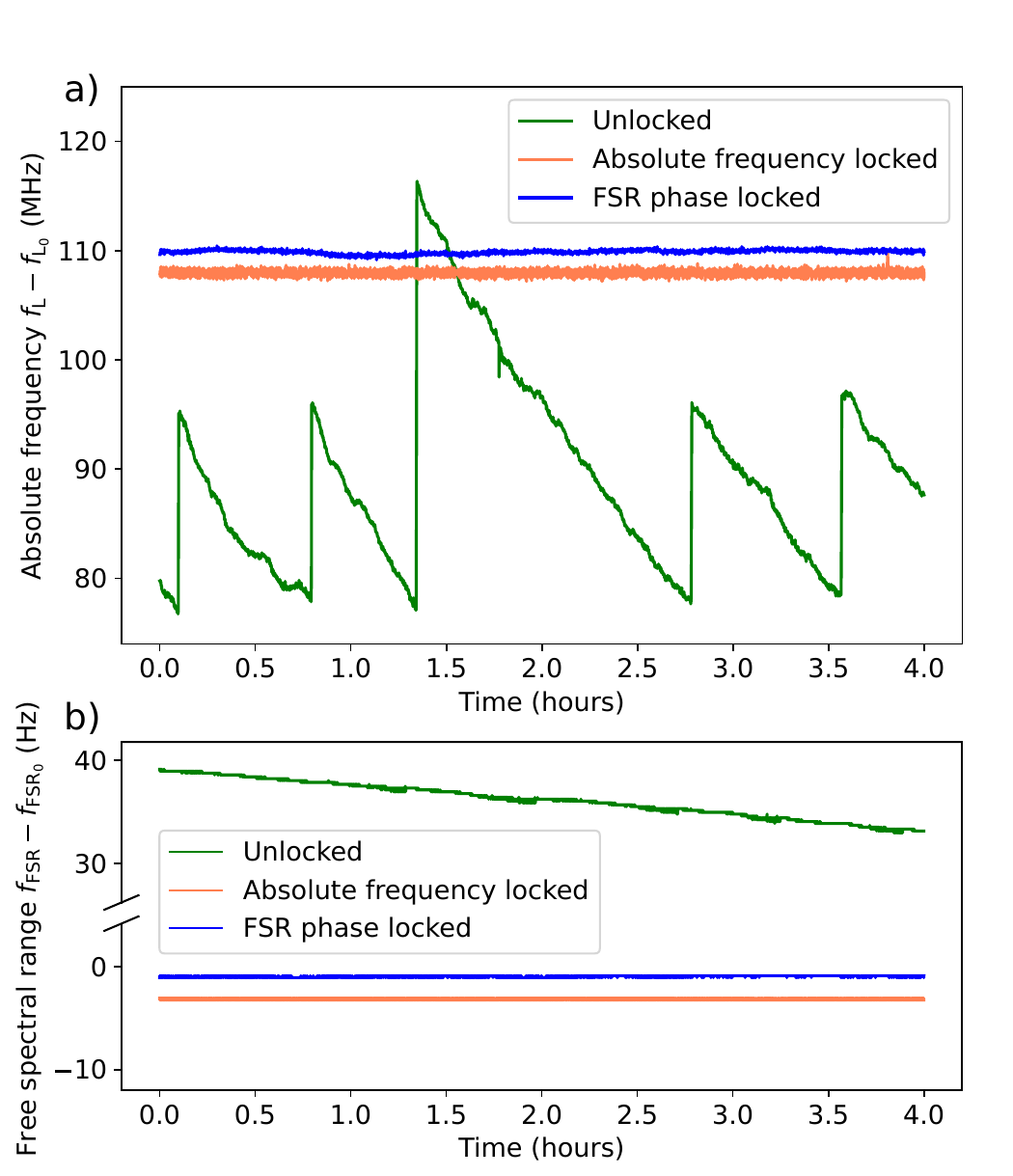}
    \caption{\label{fig:wlm} a) Absolute frequency of the ring laser $f_{\text{L}}$, shifted by $f_{\text{L}_{0}} =$ \SI{473.612}{\tera \hertz}, as measured by the wavelength meter with a rate of \SI{0.25}{\hertz} for unlocked operation, for the \textit{absolute~frequency~lock} (in-loop) and for the \textit{FSR~phase~lock} (out-of-loop). b) Beat of the free spectral range $f_{\text{FSR}}$, shifted by $f_{\text{FSR}_{0}} =$ \SI{21.4232}{\mega \hertz}, for unlocked operation, for the \textit{absolute~frequency~lock} (out-of-loop) and for the \textit{FSR~phase~lock} (in-loop).}
\end{figure}

To evaluate the performance of the \textit{absolute~frequency~lock} and the \textit{FSR~phase~lock} in comparison to an unlocked ring laser, we performed three four-hour measurements during consecutive nights, each starting at \SI{11}{pm}. The absolute lasing frequency $f_{\text{L}}$, the free spectral range $f_{\text{FSR}}$, and the Sagnac frequency $\delta f$ are logged during these measurement intervals.

\begin{figure} [t]
    \includegraphics[width=\columnwidth]{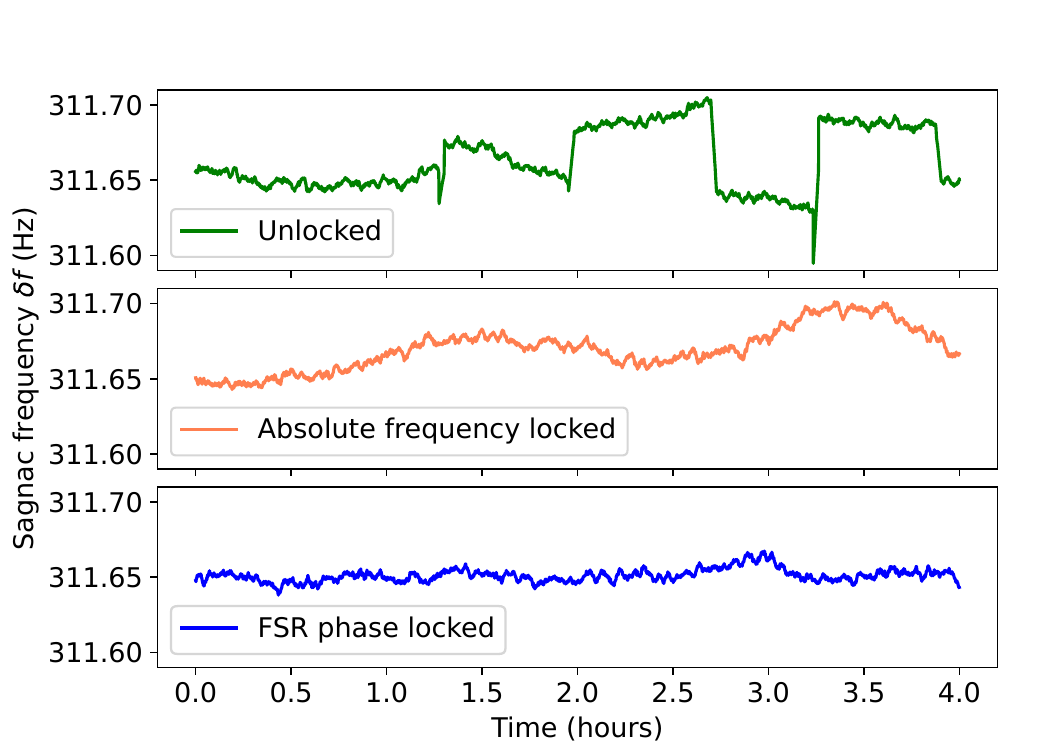}
    \caption{\label{fig:sagnac} Sagnac frequency $\delta f$, plotted with a running mean average of \SI{100}{\second} interval length.}
\end{figure}

The absolute lasing frequency was constantly measured by the wavelength meter in \SI{4}{\second} intervals, as shown in Fig.~\ref{fig:wlm}~a). Without locking, the frequency $f_{\text{L}}$ drifts rather linearly at a rate of \SI{25}{\mega \hertz \per \hour}. Furthermore, five discontinuities can be observed, which are caused by hopping of the laser mode due to drifting of the laser frequency by more than one FSR. Mode jumps by two FSR are observed as well. Both locking schemes work very well at canceling drifting and therefore also suppress mode hops. In direct comparison, the \textit{FSR~phase~lock} shows less variation with a standard deviation of $f_{\text{L}}$ of \SI{196}{\kilo \hertz}, compared to \SI{334}{\kilo \hertz} in the \textit{absolute~frequency~lock} case, caused mainly by the resolution of the WLM of only \SI{100}{\kilo \hertz}, as well as the rather slow locking bandwidth due to the long required integration time. Interpreting these deviation values as fluctuations of the $P =$ \SI{14}{\meter} perimeter yield $\Delta P=$ \SI{5.8}{\nano \meter} and \SI{9.9}{\nano \meter}, respectively, equating to relative stabilities of $4.1\times 10^{-10}$ and $7.1\times 10^{-10}$, respectively.

Furthermore, using a directional coupler, a \SI{20}{\decibel} attenuated part of the amplified and filtered $4 \cdot f_{\text{FSR}}$ beat signal is constantly monitored as the peak position in a spectrum analyzer sweep at a rate of \SI{1}{\hertz} with \SI{0.18}{\hertz} resolution, referenced to an atomic clock. Similarly to $f_{\text{L}}$, an approximately linear downward trend of \SI{1.5}{\hertz \per \hour} is observed in unlocked operation, as shown in Fig.~\ref{fig:wlm}~b). When operating the ring laser using any of the two locking schemes, the measured value of $f_{\text{FSR}}$ fluctuates between two 0.18-Hz increments of the spectrum analyzer's resolution. This fluctuation is faster in the \textit{absolute~frequency~lock} case.

A datalogger, referenced to an atomic clock, is used to digitize the Sagnac beat signal $S(t)$ at a sampling rate of \SI{7}{\kilo \hertz}, and its frequency $\delta f$ is estimated following Ref.~\cite{Igel2021} in real-time: The Hilbert transformation $H[S(t)]$ of the signal is computed, as well as the time derivatives, indicated by $d_{t}$. The instantaneous Sagnac frequency then follows according to
\begin{align}
    \label{align:hilbert}
    \delta f (t) = \frac{S(t) d_{t}H[S(t)] - H[S(t)]d_{t}S(t)}{2\pi (S(t)^{2} + H[S(t)]^{2})}.
\end{align}
It is then best viewed as a running mean average with a width of \SI{100}{\second}, as shown in Fig.~\ref{fig:sagnac}. In unlocked operation, the time series is characterized by drifts and discontinuous jumps of $\delta f$, far larger than expected from geometric changes in the scale factor. The cause of these discontinuities are changes in laser dynamics, in particular a near-instantaneous change of backscatter coupling amplitudes, which are likely to happen in a unlocked drifting system with varying lasing frequency $f_{\text{L}}$ and mode index $m$. Stabilizing the perimeter by either one of the locking schemes drastically reduces the likelihood of discontinuous jumps to the point that none were observed during this measurement. The \textit{FSR~phase~lock} reduces residual drifts in $\delta f$ the best. One origin for the residual variations in $\delta f$ is laser dynamics that can be accounted for by a backscatter correction \cite{Hurst2014} and a null-shift correction \cite{Beghi2012,DiVirgilio2019}, which are on our current agenda. Another origin is local tilt and beamwander, which will also be detected and corrected for in future improvements. 

\begin{figure} [t]
    \includegraphics[width=\columnwidth]{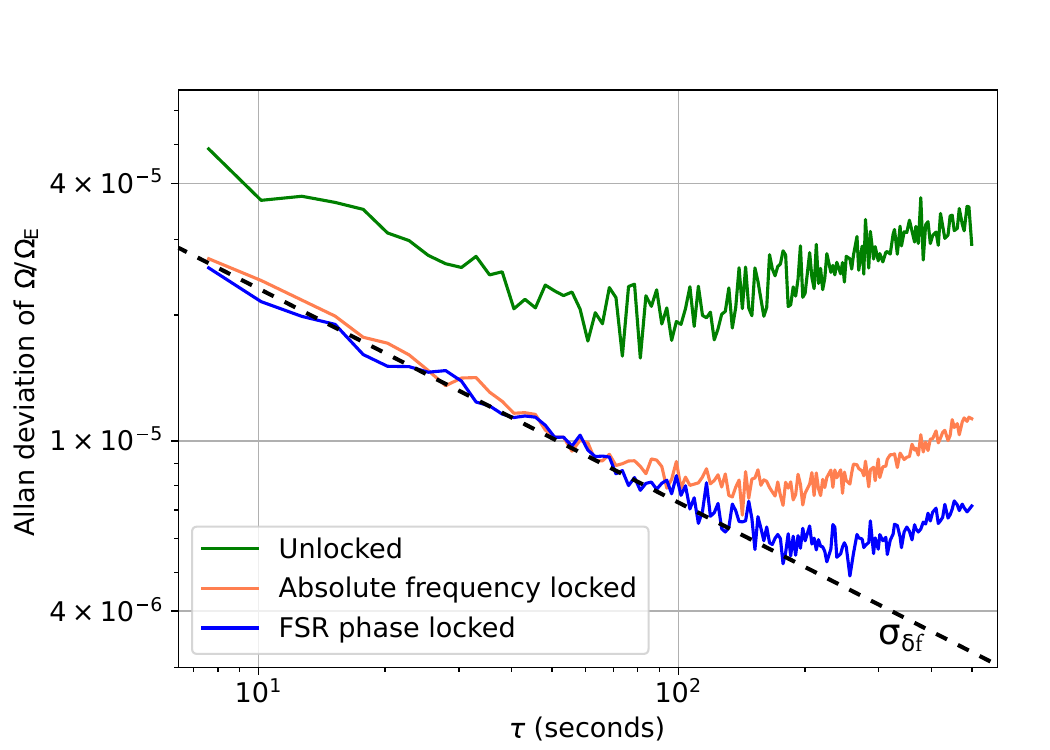}
    \caption{\label{fig:allandev} Classic Allan deviations of the measured Sagnac frequency time series shown in Fig.~\ref{fig:sagnac}. The data is normalized to Earth's rotation rate $\Omega_{\text{E}}$ according to Eq.~\ref{align:sag}. The sensitivity of $\sigma_{\delta f} =\mathrm{\SI{5.5}{\nano \radian \per \second \per \sqrt{\hertz}}}$ is marked with a dashed line.}
\end{figure}

To quantify the overall performance of the ring laser, Allan deviations of the measured Sagnac signals are calculated according to Ref.~\cite{Allan1966, Sullivan1990}, see Fig.~\ref{fig:allandev}. During \textit{FSR~phase~locked} operation, the best stability and sensitivity are observed with \SI{280}{\pico \radian \per \second} at \SI{250}{\second} integration time, corresponding to $5\times 10^{-6}\ \Omega_{\text{E}}$. In the case of both locking schemes, a sensitivity of $\sigma_{\delta f} =\mathrm{\SI{5.5}{\nano \radian \per \second \per \sqrt{\hertz}}}$ is reached, which is marked as a dashed line in Fig.~\ref{fig:allandev}. The theoretical sensitivity floor \cite{Schreiber2013}, only limited by quantum shot noise, would be $\sigma_{\delta f_{theo}} =\mathrm{\SI{0.16}{\nano \radian \per \second \per \sqrt{\hertz}}}$. Our experimental sensitivity is reduced by finite efficiency of the photodetector and white noise on the electronics.

\section{Conclusion}

We have presented two cost-effective and easy-to-implement methods that allow for the stabilization of the perimeter of a large ring laser at the level of a few parts in $10^{-10}$. Both locking methods increased the sensitivity by a factor of about two and significantly improved the stability of the Sagnac signal. While the Allan deviation reaches its minimum already at about \SI{80}{\second} in the unlocked case, this timescale is increased to \SI{150}{\second} for the absolute frequency lock and further to \SI{250}{\second} for the FSR phase lock. Acknowledging that the current best ring lasers reach a stability of order $10^{-9}$ in $\delta f$, we conclude that the performance of heterolithic ring lasers will not be limited by variations of the perimeter. In the next step, we will also stabilize the beam path using position-sensitive detectors and piezo actuators on the mirrors.

\section*{Acknowledgments}
We are indebted to U.~Hugentobler and the Forschungseinrichtung Satellitengeodäsie at TU Munich for the loan of the ring laser hardware. We acknowledge fruitful discussions with J.~Kodet, H.~Igel, and A.~Brotzer, as well as experimental support from P.~Hänisch. 

\section*{Data Availability Statement}
The data that support the findings of this study are available from the corresponding author upon reasonable request.

\section*{Disclosures}
The authors declare no conflicts of interest.

\section*{Funding}
We acknowledge financial support from the European Research Council through grant No.~101123334.



\bibliography{bib}

\end{document}